\documentclass{my}

\begin{document}

\title{Mode Regularization of the Configuration Space Path Integral
for a Particle in Curved Space}

\author{F. Bastianelli}

\address{Dipartimento  di Fisica, Universit\`a di Modena,
via Campi 213-A, I-41100 Modena, 
\\ and \\
INFN, Sezione di Bologna, via Irnerio 46, I-40126 Bologna, Italy
\\E-mail: bastiane@unimo.it}


\maketitle

\abstracts{
The proper definition and evaluation of the configuration space path integral 
for the motion of a particle in curved space is a notoriously tricky problem.
We discuss a consistent definition which makes use of an expansion in Fourier 
sine series of the particle paths.
Salient features of the regularization are the Lee-Yang ghosts fields and 
a specific effective potential to be added to the classical action.
The Lee-Yang ghost fields are introduced to exponentiate 
the non-trivial path integral measure and make the perturbative loop 
expansion finite order by order, whereas the effective potential is necessary 
to maintain the general coordinate invariance of the model. 
We also discuss a three loop computation which tests the mode 
regularization scheme and reproduces consistently the perturbative 
De Witt solution of the heat kernel.}

\section{Introduction}

The Schr{\"o}dinger equation for a particle moving in a 
curved space with metric $g_{\mu\nu}(x)$ has a wide range of applications 
going from non-relativistic diffusion problems 
(described by its euclidean version) 
to the relativistic description of particles in a curved space-time. 
However it cannot be solved exactly for an arbitrary background metric 
and one has to use some kind of perturbation theory.
A very useful perturbative solution can be obtained by employing
an ansatz introduced by De Witt\cite{one}, known as the 
heat kernel ansatz. This ansatz makes use of a power series expansion in 
the time of propagation of the particle. The coefficients of the power
series are then determined iteratively by requiring that the 
Schr{\"o}dinger equation be satisfied perturbatively.

Equivalently, the solution of the Schr{\"o}dinger equation 
should admit a path integral representation, as proposed by Feynman 
fifty years ago.
However, a proper definition of the path integral in curved space 
is not straightforward, and the history of this subject has been
quite complicated and controversial\cite{four}.
The essential complications arise mainly from: 
$i)$ the non-trivial path integral measure and
$ii)$ the proper discretization of the action necessary to 
regulate the path integral.
A quite extensive literature has been produced over the years addressing  
especially the latter point. 

In this talk we discuss a consistent method of defining the path integral 
by employing mode regularization, as it is standard in many 
calculations done in quantum field theory.
The methods extends the one already
employed by Feynman and Hibbs in discussing 
mode regularization of the path integral in flat space
and has been introduced and successively refined  in \cite{six},
\cite{seven}, \cite{eight} and \cite{eightbis},
where quantum mechanics was used mostly to compute one loop trace 
anomalies of certain quantum field theories.
The key feature is to employ ghost fields to treat the non-trivial 
path integral measure as part of the action. 
These ghost fields have been named Lee-Yang ghosts and
allow to take care of the non-trivial path integral 
measure at higher loops in a consistent manner.
The path integral is then defined by expanding all fields, including 
the ghosts, in a sine expansion about the classical trajectories 
(or any trajectory which includes the correct boundary values)
and integrating over the corresponding Fourier coefficients.
The necessary regularization is obtained by integrating
all coefficients up to a fixed mode $M$, which is eventually taken to 
infinity. A drawback of mode regularization is that it doesn't 
respect general coordinate invariance
of the target space: a particular non-covariant 
counterterm arising at two loops has to be used in order to restore 
that symmetry\cite{eight}.
General arguments based on power counting (quantum mechanics can
be thought as a super-renormalizable
field theory) plus the fact that 
the correct trace anomalies are obtained by the use of this path integral
suggest that the mode regularization described above is consistent to 
any loop order without any additional input.
We have tested this construction by a full three loop computation 
of the transition amplitude\cite{eightbis}. 
The result is found to be right since it 
solves the correct Schr{\"o}dinger equation at the required loop order.
This gives a powerful check on the method of mode regularization for 
quantum mechanical path integrals on curved space and puts it on solid 
basis. In addition, the computation can be readily  
extended and compared to the time discretization method developed in 
refs. \cite{nine}, which is also based on the use of the Lee-Yang ghosts.
That method requires its own specific counterterm to restore general 
coordinate invariance and we checked that it gives
the same correct answer.

In the next section we  present the relevant items of mode 
regularization and give the three loop transition amplitude 
in Riemann normal coordinates. 
%Details of the latter will be found in \cite{eightbis}.

\section{Mode Regularization and the 3 Loop Transition Amplitude}

The euclidean Schr{\"o}dinger equation for a particle of 
unit mass moving in a $D$-dimensional space with metric 
$g_{\mu\nu}$ and coupled to a scalar potential $V$ is 
\be
- {\partial \over \partial t} \Psi =  \bigl [- {1\over 2 } \nabla^2  + V(x)
 \bigr ] \Psi.
\label{schr} 
\ee
It can be solved perturbatively by the heat kernel ansatz of 
De Witt\cite{one}:
\be
\Psi(x,y,t) = {1 \over {(2 \pi t)^{D \over 2}}} \
{\rm e}^{ - { \sigma(x,y)\over  t}}
\sum_{n=0}^\infty a_n(x,y) t^n
\label{ansatz}
\ee
which depends parametrically on the point $y^\mu$ specifying the 
boundary condition $ \Psi(x,y,0)=  
%{\delta^D(x-y) \over \sqrt{g(x)}}
g^{-{1\over2}} \delta^D(x-y) 
$
and with $\sigma(x,y)$ the Synge world function.

On the other hand, following refs. \cite{six,seven,eight,eightbis},
we can write the transition amplitude 
for the particle to propagate from the initial point
$x^\mu_i$ at time $t_i$ to the final point
$x^\mu_f$ at time $t_f$ as follows
\bea
&&\langle x^\mu_f,t_f | x^\mu_i,t_i \rangle  =
\int_{_{x(-1) = x_i}}^{^{x(0)=x_f}} 
\! \! \! \! \! \! \! \! \! \! \! \!  \! \! \!  
{\cal D} x  {\cal D} a  {\cal D} c  {\cal D} c \
\exp \biggl [ - { 1\over {\beta }} S \biggr ]
\label{pi}
\\
&&
S =
\int_{-1}^{0}  \! \! \! 
d\tau \ \biggl [
{1\over 2} g_{\mu\nu}(x) (\dot x^\mu \dot x^\nu + a^\mu a^\nu +
b^\mu c^\nu) + \beta^2 \biggl ( V(x) + V_{MR}(x) \biggr ) \biggr ]
\label{quac}  \\
&& 
 V_{MR} = 
{1\over 8} R  -{1\over 24} g^{\mu\nu} g^{\alpha \beta} g_{\gamma\delta}
\Gamma_{\mu\alpha}{}^\gamma \Gamma_{\nu\beta}{}^\delta 
\label{ct} \\
&& 
{\cal D} x  {\cal D} a  {\cal D} c  {\cal D} c
= (2\pi \beta)^{-{D\over 2}} \prod_{m=1}^{\infty}
\prod_{\mu=1}^D m \,
d x^\mu_m  d a^\mu_m  d b^\mu_m d c^\mu_m. \label{mes}
\eea
For commodity we have shifted and rescaled the time parameter 
in the action,
$t= t_f+ \beta \tau$
with $\beta = t_f - t_i$, so that $-1 \leq \tau \leq 0$.
The total time of propagation 
$\beta$ plays the role of the Planck constant 
$\hbar$ (which we have set to one) and counts the number of loops.  
In the loop expansion generated by $\beta$ the potentials 
$V$ and $V_{MR}$ start contributing only at two loops.
The full action $S$ includes terms proportional to the
Lee-Yang ghosts, namely the commuting ghosts $a^\mu$ and the 
anticommuting ghosts  $b^\mu$ and $ c^\mu$.
Their effect is to reproduce a formally covariant measure.
The regulated measure is obtained by expanding all fields in a sine series 
around a background value
\be
\phi^\mu(\tau) = \phi_{bg}^\mu(\tau) + \phi_{qu}^\mu(\tau) ; \ \ \ 
\phi_{qu}^\mu(\tau) =\sum_{m=1}^{\infty} \phi^\mu_m \sin (\pi  m \tau)
\ee
where $\phi =( x^\mu, a^\mu, b^\mu, c^\mu) $ and 
$\phi_{bg} =( x^\mu_{bg}, 0,0,0)$,
and defining it as product of the standard measures for 
the coefficients $\phi^\mu_m$ as in (\ref{mes}).
The regularization is obtained by integrating up to a fixed mode 
number $M$ that eventually is taken to infinity.
This regularization does not respect  general coordinate invariance
which is recovered only thanks to the effect of the effective potential
$V_{MR}$ in (\ref{ct}). 
This way the path integral (\ref{pi}) solves eq. (\ref{schr}). 
In fact, one can start computing it by using the standard perturbative 
expansion and using Riemann normal coordinates 
centred at $x_f^\mu$. 
Ambiguities in Feynman diagrams are resolved by mode regularization. 
Thus one gets the transition amplitude at 3 loops 
which is shown to solves eq. (\ref{schr}) at the required loop 
order\cite{eightbis}. The result can be cast in a manifestly 
symmetric form by using symmetrized
quantities $\overline{ A} \equiv {1\over 2} [A(x_i)  + A (x_f)] $
and with  $\xi^\mu \equiv x^\mu_i- x^\mu_f$ 
considered of order $\beta^{1\over2}$
\bea
&& \langle x^\mu_f,t_f | x^\mu_i,t_i \rangle  
= {1\over (2 \pi \beta)^{{D\over 2}}}
\exp \biggl [ 
-{1\over 2 \beta }\xi^\mu\xi^\nu \overline{ g_{\mu\nu}}
-{1\over 12} \xi^\alpha \xi^\beta \overline{R_{\alpha\beta}}
-\beta \biggl ({1\over 12} \overline{ R} + \overline{ V }\biggr )
\nonumber
\\
&&
+ \xi^\alpha \xi^\beta \xi^\gamma \xi^\delta 
\biggl (
{1\over 360} \overline{ R_{\alpha\mu\nu \beta} R_\gamma{}^{\mu\nu}{}_\delta} +
{1\over 120} \overline{\nabla_\gamma \nabla_\delta R_{\alpha\beta}}
\biggr ) 
\nonumber
\\
&&
+\beta \xi^\alpha \xi^\beta 
\biggl ( 
{1\over 360} \overline{ R_{\alpha\mu\nu\lambda} R_\beta{}^{\mu\nu\lambda}} 
-{1\over 720} \overline{  R_{\alpha\mu\nu\beta} R^{\mu\nu}}
-{1\over 720} \overline{  R_{\alpha\mu} R_\beta{}^\mu}
\nonumber
\\
&& 
-{1\over 240} \overline{ \nabla^\mu \nabla^\nu R_{\alpha\mu\nu\beta}}
+{1\over 160} \overline{ \nabla_\alpha \nabla_\beta R} 
+{1\over 12} \overline{ \nabla_\alpha \nabla_\beta V} 
\biggr ) 
\nonumber
\\
&&
+ \beta^2 
\biggl ( 
{1\over 720} \overline{ R_{\alpha\mu\nu\beta}^2} 
- {1\over 720} \overline{ R_{\alpha\beta}^2}
- {1\over 120} \overline{ \nabla^2 R} -{1\over 12} \overline{ \nabla^2 V}
\biggr ) + O(\beta^{5\over 2})
\biggr].
\label{resdue}
\eea
By comparing this result with the expression in (\ref{ansatz})
one can extract the leading terms of the coefficients
$a_0,a_1,a_2$ for non-coinciding points.

\section*{Acknowledgements}
I wish to thank  O. Corradini, K. Schalm and  P. van Nieuwenhuizen
for their invaluable collaboration at various stages of this research.

\end{document}